\documentclass[twocolumn,english,aps,pra,twocolum,superscriptaddress,natbib,amsmath,amssymb,floatfix,superscriptaddress,footinbib]{revtex4-2}
\usepackage[colorlinks=true,citecolor=blue,linkcolor=magenta]{hyperref}
\usepackage{color} 

\usepackage[addedmarkup=colored, authormarkupposition=left]{changes} 

\usepackage{soul}
\definechangesauthor[name={TJK}, color=red]{TJK}
\definecolor{hugoColor}{RGB}{59,134,255}
\definechangesauthor[name={HL}, color=hugoColor]{HL}

\usepackage[english]{babel}
\usepackage{amsmath,amsfonts,amssymb}
\usepackage[T1]{fontenc}
\usepackage{xurl}

\usepackage{amsmath}
\usepackage{siunitx}
\usepackage[version=4]{mhchem}

\DeclareSIUnit \baud {Bd}

\usepackage{amsfonts}
\usepackage{amssymb}

\usepackage{epstopdf}
\usepackage{graphicx}
\usepackage{float}
\graphicspath{{./Figures/}}
\usepackage[utf8]{inputenc}
\UseRawInputEncoding

\newcommand{\fref}[1]{Fig.~\ref{#1}}
\newcommand{\SiN}[0]{$\mathrm{Si}_3\mathrm{N}_4$}
\newcommand{\LN}[0]{$\mathrm{LiNbO}_3$} 
\newcommand{\LT}[0]{$\mathrm{LiTaO}_3$}

\newcommand{\threefive}[0]{\uppercase\expandafter{\romannumeral3}-\uppercase\expandafter{\romannumeral5}~}

\begin{document}

    \title{Heterogeneously integrated lithium tantalate-on-silicon nitride modulators for high-speed communications}
    
    \author{Jiachen~Cai}\thanks{These authors contributed equally.}
    \affiliation{State Key Laboratory of Materials for Integrated Circuits, Shanghai Institute of Microsystem and Information Technology, Chinese Academy of Sciences, Shanghai, China}
    \affiliation{Institute of Physics, Swiss Federal Institute of Technology Lausanne (EPFL), CH-1015 Lausanne, Switzerland}
    
    \author{Alexander~Kotz}\thanks{These authors contributed equally.}
    \affiliation{Institute of Photonics and Quantum Electronics (IPQ), Karlsruhe Institute of Technology (KIT), 76131 Karlsruhe, Germany}

    \author{Hugo~Larocque}\thanks{These authors contributed equally.}
    \affiliation{Institute of Physics, Swiss Federal Institute of Technology Lausanne (EPFL), CH-1015 Lausanne, Switzerland}
    
    \author{Chengli~Wang}
    \affiliation{State Key Laboratory of Materials for Integrated Circuits, Shanghai Institute of Microsystem and Information Technology, Chinese Academy of Sciences, Shanghai, China}
    \affiliation{Institute of Physics, Swiss Federal Institute of Technology Lausanne (EPFL), CH-1015 Lausanne, Switzerland}

    \author{Xinru~Ji}
    \affiliation{Institute of Physics, Swiss Federal Institute of Technology Lausanne (EPFL), CH-1015 Lausanne, Switzerland}

    \author{Junyin~Zhang}
    \affiliation{Institute of Physics, Swiss Federal Institute of Technology Lausanne (EPFL), CH-1015 Lausanne, Switzerland}

    \author{Daniel~Drayss}
    \affiliation{Institute of Photonics and Quantum Electronics (IPQ), Karlsruhe Institute of Technology (KIT), 76131 Karlsruhe, Germany}

    \author{Xin~Ou}
    \email[]{ouxin@mail.sim.ac.cn}
    \affiliation{State Key Laboratory of Materials for Integrated Circuits, Shanghai Institute of Microsystem and Information Technology, Chinese Academy of Sciences, Shanghai, China}

    \author{Christian~Koos}
    \email[]{christian.koos@kit.edu}
    \affiliation{Institute of Photonics and Quantum Electronics (IPQ), Karlsruhe Institute of Technology (KIT), 76131 Karlsruhe, Germany}
    
    \author{Tobias~J.~Kippenberg}
    \email[]{tobias.kippenberg@epfl.ch}
    \affiliation{Institute of Physics, Swiss Federal Institute of Technology Lausanne (EPFL), CH-1015 Lausanne, Switzerland}
    \affiliation{Institute of Electrical and Micro engineering, Swiss Federal Institute of Technology, Lausanne (EPFL), CH-1015 Lausanne, Switzerland}

\maketitle
\noindent 


    \noindent \textbf{Driven by the prospects of higher bandwidths for optical interconnects, integrated modulators involving materials beyond those available in silicon manufacturing increasingly rely on the Pockels effect.
    For instance, wafer-scale bonding of lithium niobate films onto ultralow loss silicon nitride photonic integrated circuits provides heterogeneous integrated devices with low modulation voltages operating at higher speeds than silicon photonics. However, in spite of its excellent electro-optic modulation capabilities, lithium niobate suffers from drawbacks such as  birefringence and long-term bias instability. Among other available electro-optic materials, lithium tantalate can overcome these shortcomings with its comparable electro-optic coefficient, significantly improved photostability, low birefringence, higher optical damage threshold, and enhanced DC bias stability. Here, we demonstrate wafer-scale heterogeneous integration of lithium tantalate films on low-loss silicon nitride photonic integrated circuits. With this hybrid platform, we implement modulators that combine the ultralow optical loss ($\sim$ 14.2 dB/m), mature processing and wide transparency of silicon nitride waveguides with the ultrafast electro-optic response of thin-film lithium tantalate. The resulting devices achieve a 6~V half-wave voltage, and support modulation bandwidths of up to 100~GHz.  We use single intensity modulators and in-phase/quadrature (IQ) modulators to transmit PAM4 and 16-QAM signals reaching up to 333 and 581 Gbit/second net data rates, respectively. Our results demonstrate that lithium tantalate is a viable approach to broadband photonics sustaining extended optical propagation, which can uniquely contribute to technologies such as RF photonics, interconnects, and analog signal processors.}
 
    \section{Introduction} 
    Photonic integrated circuits (PICs) provide a unique approach to scaling standardized optoelectronic technologies.~\cite{Li:21,liu:22}.
    Among PIC platforms, silicon nitride-based devices offer a range of features traditionally leveraged in optical fibers such as low propagation losses, large power handling~\cite{Liu:22_OL}, and a wide bandgap enabling transparency across a wide range of optical wavelengths~\cite{Lu:19, Lu:20, blumenthal_photonic_2020}.
    Thicker nitride waveguides, e.g. manufactured with a photonic Damascene process, additionally provide strong optical confinement and easily-achievable  anomalous group velocity dispersion~\cite{Pfeiffer:18,liu_high-yield_2021, ji_compact_2022, ji2025copper}. These features become crucial while harnessing this material's optical nonlinearities~\cite{Pfeiffer:16} and have thus lead to demonstrations pertaining to Kerr frequency combs~\cite{Pfeiffer:16, moille_parametrically_2024, moille_versatile_2025, ji2025copper}, optical frequency conversion~\cite{lu_efficient_2019, lu_efficient_2021}, travelling-wave optical parametric amplifier~\cite{riemensberger2022photonic,zhao2025ultra}, and quantum information science~\cite{lu_chip-integrated_2019, singh_quantum_2019, Vaidya:20, Aghaee:25}.
    Further deployment of this platform in applications such as optical communications and microwave photonics~\cite{kudelin_photonic_2024,He:24} requires electro-optic modulation, which silicon nitride cannot directly provide due to its amorphous nature.
    Ferroelectric-based integrated circuits readily provide such features~\cite{wang_integrated_2018, li2020lithium, Larocque:24}, and can thus introduce electro-optic modulation capabilities into silicon nitride circuits through the heterogeneous integration of materials like lithium niboate~\cite{churaev2023heterogeneously, snigirev2023ultrafast}. Improved metrics in other ferroelectrics such as lithium tantalate, which offer lower birefringence and photorefractive effects~\cite{wang_lithium_2024, wang_ultrabroadband_2024, lin2025copper}, motivate applying similar integration methods to alternative Pockels materials. Improved economies of scale for lithium tantalate substrates due to their role in 5G/6G RF filters~\cite{yan_wafer-scale_2019} further encourages the use of this material in integrated modulators.
    Here, we develop a wafer-scale bonding process for lithium tantalate on silicon nitride PICs. With this platform, we demonstrate modulation defined by a $V_\pi L = 4.08$~V$\cdot$cm and a 3-dB bandwidth close to 100 GHz in a push-pull Mach-Zehnder modulator (MZM). 
    These results enable net data transmission rates of 333 Gbit/s when operating the modulators in pulse amplitude modulation (PAM) schemes. We further extend this data transmission demonstration to \SiN-\LT~IQ modulators, thereby achieving net data transmission rates of 581 Gbit/s using quadrature amplitude modulation (QAM) signals.
    With performance metrics that can already compete with those of monolithic lithium-tantalate-on-insulator (LTOI) PICs~\cite{wang_ultrabroadband_2024, lin2025copper}, these results show that \SiN-\LT~devices can implement systems combining mature \SiN~PICs with next-generation ferroelectric thin films, thereby enabling both low-loss strong optical confinement with GHz-rate modulation.

        \begin{figure*}[htbp!]
		\centering
            \includegraphics[width=1\linewidth]{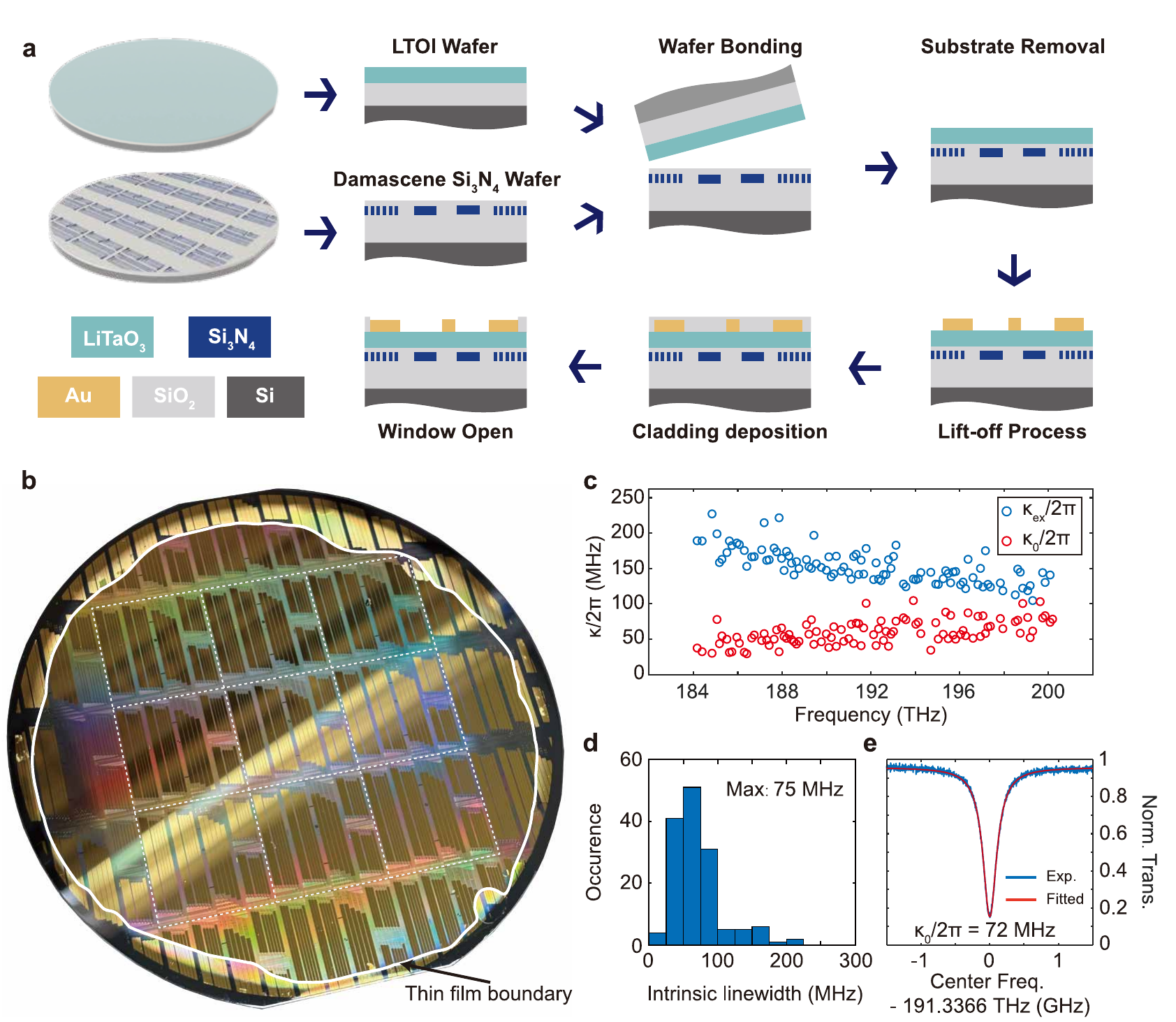}
		\caption{\textbf{Wafer-scale manufacturing of {Si\textsubscript{3}N\textsubscript{4}-LiTaO\textsubscript{3}} photonic integrated circuits.}
                (a) Schematic diagram of the main steps in the fabrication of heterogeneously integrated  {\SiN-\LT} photonic circuits. (b) Photograph showing the 100~mm {\SiN-\LT} wafer with high fabrication yield. 
                (c) Fitted value of external linewidth, $\kappa_\mathrm{ex}$, and intrinsic linewidth, $\kappa_{0}$, for a ring resonator with a \SI{2}{\micro\meter} waveguide width and 112 GHz free spectral range. (d) Corresponding fitted $\kappa_{0}$ histogram. (e) Representative normalized resonance with an intrinsic linewidth of 72 MHz.
			}
		\label{fig1}
	\end{figure*}
    
    \section{Results} 
    \noindent\textbf{Hybrid {Si\textsubscript{3}N\textsubscript{4}-LiTaO\textsubscript{3}}~PIC fabrication.} Figure~\ref{fig1}(a) outlines the process flow for the hybrid PICs, which starts with the fabrication of Si$_3$N$_4$ waveguide structures using the photonic Damascene process \cite{ji_compact_2022}. As detailed in Supplementary Section 1, the process flow begins with preform formation on a 100 mm-diameter silicon wafer with \SI{4}{\micro\meter} thick wet thermal oxide by means of deep ultraviolet (DUV) stepper lithography, dry etching and reflow under high temperatures. Low-pressure chemical vapor deposition (LPCVD) then deposits a layer of Si$_3$N$_4$, which subsequently undergoes chemical-mechanical polishing (CMP), oxide interlayer deposition, and annealing. The Si$_3$N$_4$ photonic Damascene process is free of crack formation in the highly tensile LPCVD Si$_3$N$_4$ film and provides high fabrication yields and ultralow optical propagation losses of $\mathcal{O}$(dB/m)~\cite{liu_high-yield_2021}. The patterned sample supports strip waveguides with a \SI{1}{\micro\meter} width and a \SI{0.5}{\micro\meter} height. In addition, inverse nanotapers allow for efficient edge coupling between the chip's waveguides and lensed fibers.  

    Our heterogeneous integration approach additionally involves 100~mm LTOI wafers fabricated by a hydrogen-based ion-slicing technique \cite{yan_wafer-scale_2019, wang_lithium_2024}. The resulting wafer stack consists of a 300~nm x-cut lithium tantalate thin film, a \SI{2}{\micro\meter} buried oxide layer and a \SI{525}{\micro\meter} thick silicon substrate. Following surface preparation relying on cleaning and plasma activation, the \SiN~and LTOI wafers undergo hydrophilic pre-bonding at room temperature, where van der Waals interactions establish a preliminary bond between the two wafers. A subsequent 300$^\circ$C thermal annealing process then enhances the bond strength by accelerating the polymerization of silanol (Si-OH) groups to covalent bonds (Si-O-Si) at the interface. 

    After bonding, the process continues by thinning the \LT-based donor substrate. 
    Backside wafer grinding followed by a tetramethylammonium hydroxide (TMAH) wet etch entirely removes the substrate's silicon. A buffered hydrofluoric acid (BHF) solution then eliminates the thermal oxide layer, thereby exposing the \LT~layer and forming a well-confined hybrid optical waveguide. 
    This sequence entirely lacks plasma-based processing traditionally used in \LT~nanofabrication, which for instance includes oxygen-based diamond-like carbon hardmask etching~\cite{li2023high} and argon dry etching \cite{kassabov1988argon}. With such measures, we avoid introducing  potential plasma-induced charges at the \SiN~PIC wafer's Si/SiO$_{2}$ interface, which can lead to parasitic surface conduction RF losses~\cite{shen_parasitic_2024}.
    To add high-speed coplanar waveguide (CPW) and heater electrodes to the sample, lithography based on a maskless aligner (Heidelberg Instruments MLA150) first defines the electrode layout. Thermal evaporation of 10~nm of titanium and 800~nm of gold followed by lift-off then completes the electrode's fabrication.
    Argon-based ion beam etching subsequently patterns the \LT~thin film to form adiabatic taper transitions between \SiN~waveguides and the hybrid \SiN-\LT~waveguides. The process also fully removes the \LT~near the chip facets to improve edge coupling efficiency to the hybrid \SiN-\LT~PIC. After cladding deposition, an additional exposure step followed by hydrofluoric wet etching selectively expose portions of the sample for landing probes on the manufactured electrodes. As discussed in Supplementary Section 2, these cladding openings also match the group velocities of the CPW RF field and optical waveguide mode.
    
    Figure~\ref{fig1}(b) shows a photograph of a finalized 100-mm {hybrid \SiN-\LT} wafer, where a distinct boundary line delineates the contour of the bonded \LT~film. The central nine stepper fields are fully-covered by the film, thereby highlighting the excellent fabrication consistency and large-scale yield of the wafer bonding technique. 
    
    To characterize the optical properties of the wafer's hybrid photonic structures, frequency-comb-assisted spectroscopy~\cite{del2009frequency} relying on three external-cavity diode lasers probed optical waveguide loss in fabricated {hybrid \SiN-\LT} microring resonators. Figures~\ref{fig1}(c,d) show the derived external, $\kappa_\mathrm{ex}$, and intrinsic, $\kappa_{0}$, loss rates of the ring's TE polarization resonances near the C telecommunication band. Here, $\kappa_{0}$ increases with optical frequency, which is consistent with the behavior of similar structures implemented in \LN-on-\SiN~heterogeneous PICs~\cite{snigirev2023ultrafast,churaev2023heterogeneously}. In such structures, higher optical frequencies result in a smaller mode field defined by a stronger overlap with the higher-index ferroelectric slab, thus inducing additional bending losses in the rings. Over the frequency range considered in Fig.~\ref{fig1}(d), finite element method simulations suggest the lithium tantalate film holds roughly 48\% of the mode's energy. Figure~\ref{fig1}(e) shows a typical normalized resonance in the C-band, which indicates a fitted intrinsic linewidth of 72 MHz and a corresponding propagation optical loss of $\alpha \approx 14.2$~dB/m.

\begin{figure*}[htbp!]
		\centering
        \includegraphics[width=1\linewidth]{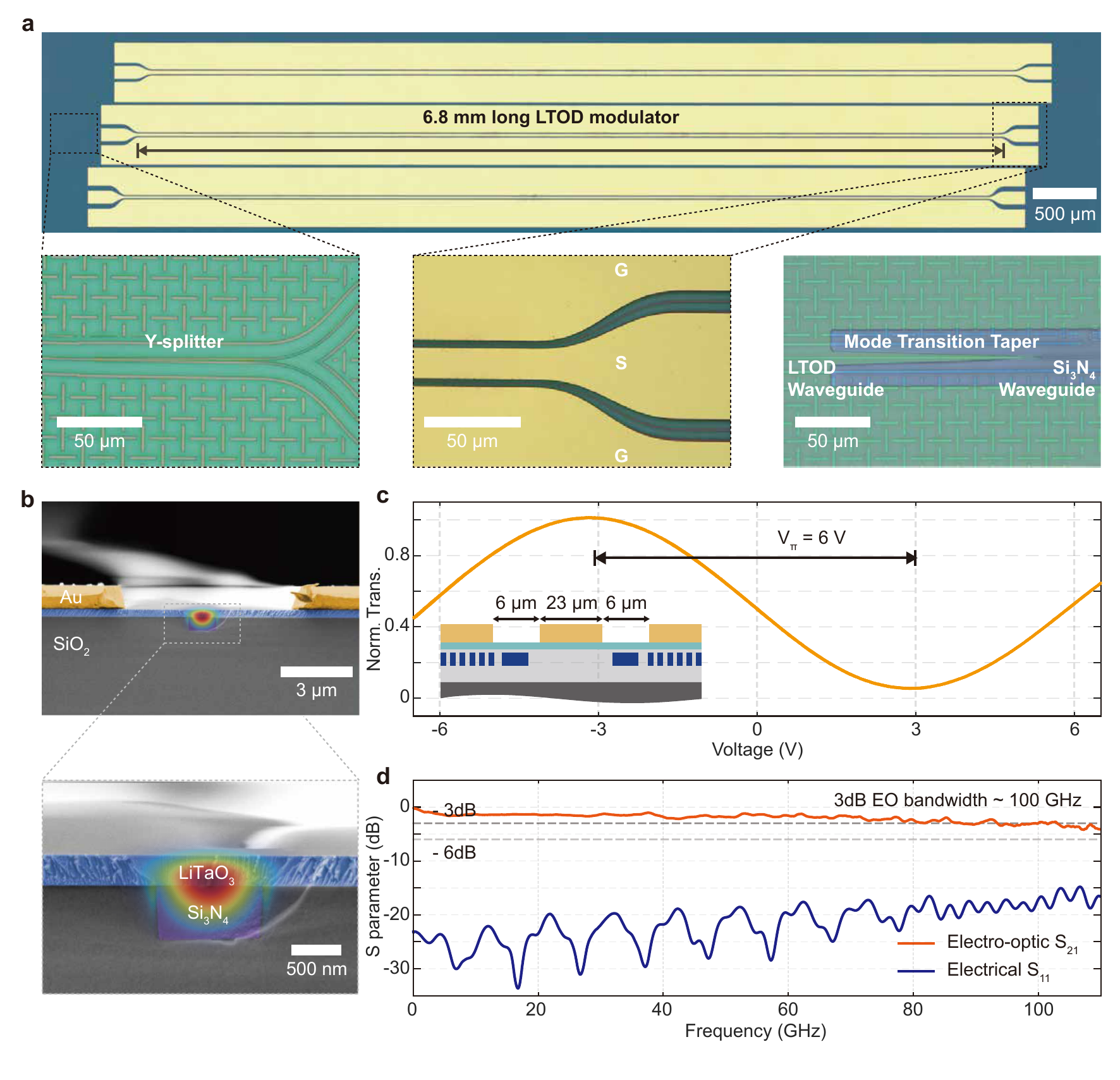}
		\caption{\textbf{{Hybrid Si\textsubscript{3}N\textsubscript{4}-LiTaO\textsubscript{3}} electro-optic modulators.}
            (a) Optical micrograph of a fabricated 6.8 mm-long hybrid modulator. Insets: optical micrographs of the modulator's underlying components, which include Y-splitters, a coplanar waveguide electrode, and tapered transitions between \SiN~waveguides and  {\SiN-\LT} waveguides with a simulated 0.28~dB insertion loss. (b) False-colored scanning electron microscopy (SEM) image of the cross-section of a manufactured {\SiN-\LT} modulator. 
            (c) Normalized transmission of a 6.8 mm-long push-pull MZM versus applied voltage. The inset diagram shows the high-speed electrode geometry with a \SI{23}{\micro\meter} signal width and a \SI{6}{\micro\meter} electrode gap. 
            (d) Measured electro-optic response (electro-optic $\mathrm{S}_{21}$) and microwave return loss (electrical reflection $\mathrm{S}_{11}$), revealing a high 3-dB bandwidth near 100 GHz.
            }
		\label{fig2}
	\end{figure*}

    \begin{figure*}[htbp!]
		\centering
		\includegraphics[width=1\linewidth]{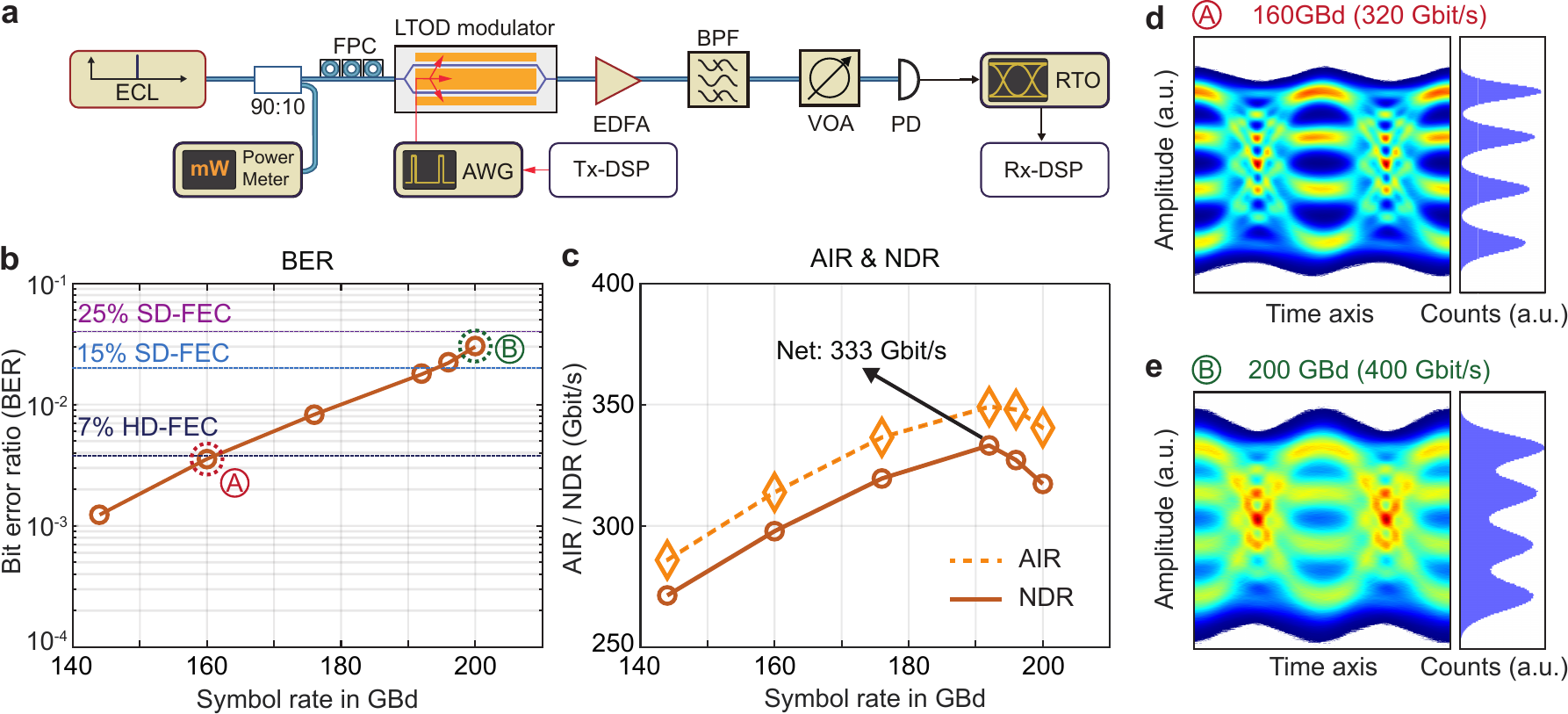}
		\caption{\textbf{Transmission experiments using intensity-modulation and direct detection (IMDD)}
			(a)~Experimental setup. ECL: External cavity laser; FPC: Fiber polarization controller; AWG: Arbitrary-waveform generator; Tx-DSP: Transmitter-digital signal processing (offline); EDFA: Erbium-doped fiber amplifier; BPF: Bandpass filter; VOA: Variable optical attenuator; PD: Photodiode; RTO: High-speed real-time oscilloscope (256 GSa/s, 105 GHz); Rx-DSP: Receiver-digital signal processing (offline). (b)~Measured bit error ratio (BER) versus symbol rate for PAM4 transmission, with horizontal dashed lines indicating the thresholds for soft-decision forward error correction with \SI{15}{\percent} and \SI{25}{\percent} coding overhead (\SI{15}{\percent} SD-FEC, \SI{25}{\percent} SD-FEC) and for hard-decision FEC with \SI{7}{\percent} coding overhead (\SI{7}{\percent} HD-FEC). The eye diagrams of two selected data points \textcircled{A} and \textcircled{B} are shown in panels (d) and (e). (c)~Achievable information rates (AIR, orange dashed line) and corresponding net data rates (NDR, solid brown line) of the IMDD measurements, showing that the highest achieved NDR is 333~Gbit/s using a PAM4 signals at a symbol rate of \SI{192}{\giga\baud}. (d), (e)~Eye diagrams (left) at selected symbol rate of \SI{160}{\giga\baud} PAM4 (d) and \SI{200}{\giga\baud} PAM4 (e), labeled by \textcircled{A} and \textcircled{B} in panel (b), along with the associated histograms (right) of the reconstructed signal amplitudes in the center of the corresponding symbol slot.
            }
		\label{fig3}
	\end{figure*}

    \noindent\textbf{Electro-optic performance.} To demonstrate electro-optic capabilities in our hybrid {\SiN-\LT} platform, we rely on electro-optic modulators consisting of 6.8~mm-long MZM pairs operating in a push-pull configuration with high-speed CPW electrodes. Figure~\ref{fig2}(a) shows a micrograph of such a modulator. Supplementary Section 3 provides simulated data regarding the transmission of some of its underlying components. The CPW design features a signal-to-ground spacing that accounts for trade-offs between metal-induced optical losses and modulation. From the cross-sectional SEM images in Figs.~\ref{fig2}(b) and the inset diagram in Figs.~\ref{fig2}(c), our CPW design balances out these two factors with a \SI{23}{\micro\meter} signal electrode width and a \SI{6}{\micro\meter} gap separating the ground and signal electrodes. Simulations presented in Supplementary Section 2 suggest this configuration yields metal spacing-induced propagation losses of 0.1~dB/cm and a voltage length product of $V_{\pi}L\sim 5.5 $~V$\cdot$cm. As further discussed in the Supplementary Information, the chosen signal width also affects the modulator's underlying bandwidth by altering RF propagation losses and velocity mismatching between the modulator's co-propagating RF and optical fields. 
    
    We first characterize the modulator's electro-optic performance under quasi-DC operation. Coupling 1550~nm continuous-wave light into the {\SiN-\LT} MZM exhibited a fiber-to-chip coupling loss of approximately 5~dB. Figure~\ref{fig2}(c) shows the measured output power while applying a 100~Hz triangle voltage wave via an RF probe. These results indicate a $V_{\pi}=6$~V half-wave voltage for a 6.8~mm long push-pull configuration, which corresponds to a $V_{\pi}L=4.08 $~V$\cdot$cm voltage-length product. Additional data provided in Supplementary Section 4 suggest this response is consitent down to modulation frequencies of at least 1~Hz and over optical wavelengths ranging from \SI{1500}{\nano\meter} to \SI{1630}{\nano\meter}. As highlighted in \fref{fig2}(d), probing the modulator's response with a vector network analyzer (VNA) provided the modulator's bandwidth. The resulting $\mathrm{S}_{21}$ electro-optic response exhibits excellent flatness within 3~dB across frequencies ranging from 25 MHz to 110 GHz while keeping $\mathrm{S}_{11}$ microwave reflections below $-15$~dB. These metrics reach levels comparable to state-of-the-art figures achieved in other electro-optic PIC platforms~\cite{wang_ultrabroadband_2024,xu2020high,han2023slow}, which underscores the significant potential of {\SiN-\LT} for applications requiring high-speed electro-optic modulation.

   \noindent\textbf{Data transmission experiments.} To demonstrate the viability and performance of our {\SiN-\LT} platform, we use our devices in high-speed optical data communication experiments, covering both intensity-modulation and direct-detection (IMDD) and coherent modulation schemes. Figure~\ref{fig3}(a) shows the setup for the IMDD experiment. Herein, a tunable external-cavity laser (ECL) operating in the C-band provides the optical carrier. A fiber-based polarization controller (FPC) then adjusts the output polarization before coupling to the quasi-TE mode, having a dominant electric field parallel to the substrate plane, of the {\SiN-\LT} MZM via a pair of lensed fibers. We drive the MZM with electrical PAM4 signals, that are synthesized by offline digital signal processing (Tx-DSP) based on pseudo-random bit sequences (PRBS) and root-raised cosine (RRC) pulse-shaping filters~\cite{wang_ultrabroadband_2024}, and that are converted to the analogue domain using a high-speed arbitrary waveform generator (AWG, M8199B, Keysight Technologies Inc.). 20 cm-long RF cables send the drive signals to the coplanar transmission line of the MZM via a first impedance-matched probe in a ground-signal-ground configuration. A second probe terminates the transmission line with a \SI{50}{\ohm} coaxial termination. Tx-DSP compensates for frequency-dependent RF loss up to the input of the feeding probe by implementing a linear minimum-mean-square-error predistortion. An erbium-doped fiber amplifier (EDFA) brings the rather weak output signal of -10~dBm from the MZM to a power level near 10~dBm compatible with the receiver. This receiver comprises a high-speed photodiode (PD, Finisar Corp.), which is directly connected to a high-speed real-time oscilloscope (RTO, UXR 1004A, Keysight Technologies Inc.) with a sampling rate of 256 GSa/s and an analogue bandwidth of 105 GHz. The EDFA is followed by an optical bandpass filter (BPF) to suppress out-of-band amplified spontaneous-emission (ASE) noise and by a variable optical attenuator (VOA, LTB-1, EXFO Inc.) that adjusts the optical power to 10 dBm, the maximum input power accepted by the photodiode. At the receiver, we use offline digital signal processing (Rx-DSP) to extract and demodulate the PAM4 data. As further elaborated in Supplementary Section 5, the Rx-DSP suite comprises standard algorithms such as timing recovery, linear Sato equalization, and an additional decision-directed least-mean-square equalizer.

   In our experiment, we used this setup to operate the device at different PAM4 symbol rates between 144 GBd and 200 GBd. Figure~\ref{fig3}(b) provides the resulting bit error ratios (BER) along with the BER thresholds for different forward error correction (FEC) schemes indicated by dashed horizontal lines. For symbol rates of 160 GBd (line rate 320 Gbit/s) and lower, the BER is below the threshold for hard-decision FEC with \SI{7}{\percent} coding overhead (\SI{7}{\percent} HD-FEC)~\cite{itu_g.975.1}, and it stays below the threshold for soft-decision FEC (SD-FEC) with \SI{15}{\percent} coding overhead~\cite[Table~7.5]{GraelliAmat2020} up to symbol rates of 192 GBd (line rate 384 Gbit/s). For 200 GBd (line rate 400 Gbit/s), we find a BER value that is still compatible with soft-decision forward-error correction (SD-FEC) with \SI{25}{\percent} coding overhead (\SI{25}{\percent} SD-FEC)~\cite[Table~7.5]{GraelliAmat2020}. Figure~\ref{fig3}(d) and \fref{fig3}(e) depict the reconstructed eye diagrams of the 160 GBd and the 200 GBd signals, respectively, along with the histograms taken at the center of the symbol slot. The data points corresponding to these eye diagrams are marked \textcircled{A} and \textcircled{B} in \fref{fig3}(b). While the signal quality clearly leaves room for improvement, the demonstrated symbol rates and line rates can already compete with those achieved by standalone lithium-tantalate-on-insulator (LTOI) MZM~\cite{wang_ultrabroadband_2024}.


  \begin{figure*}[htbp!]
		\centering
            \includegraphics[width=1\linewidth]{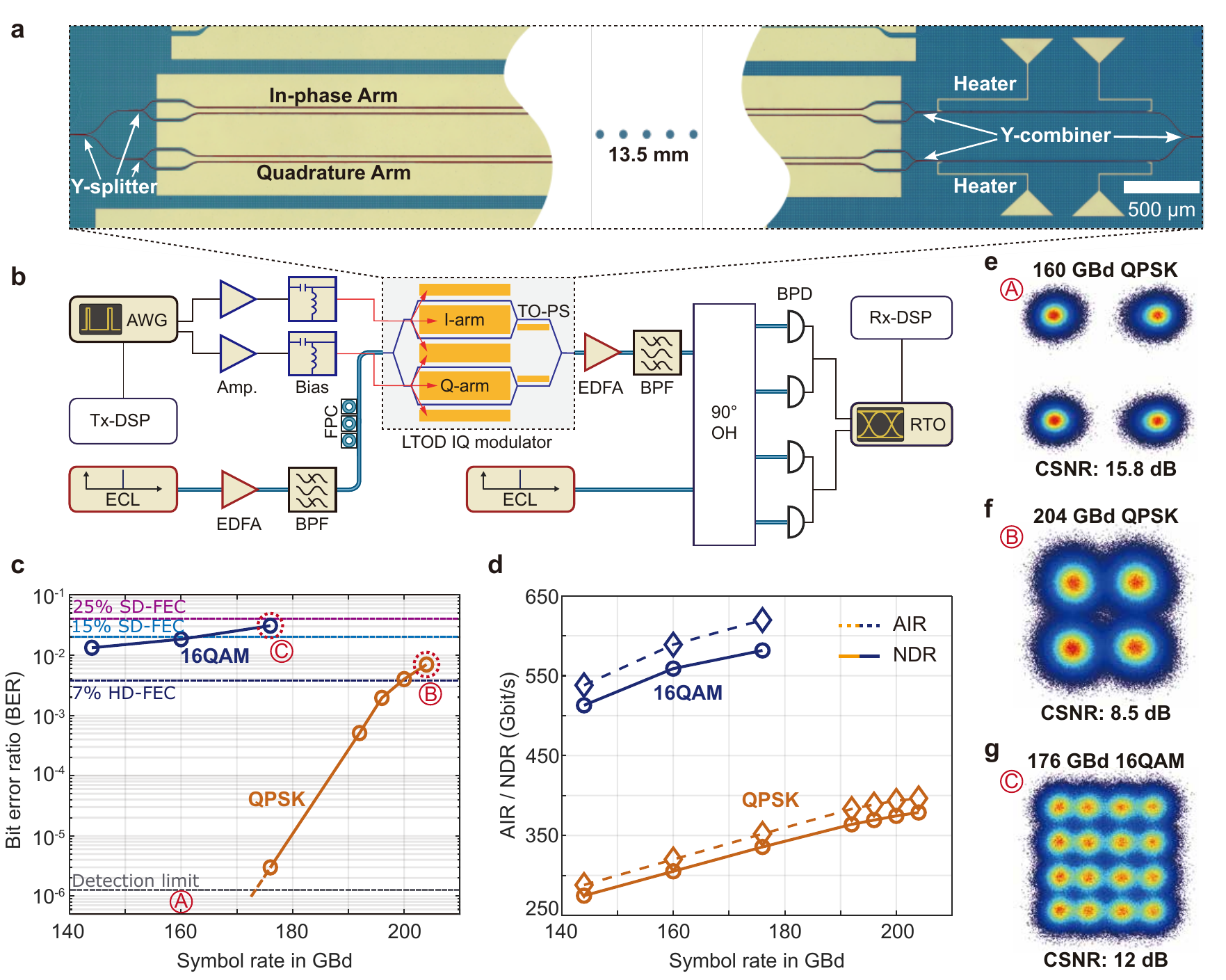}
		\caption{\textbf{{Hybrid Si\textsubscript{3}N\textsubscript{4}-LiTaO\textsubscript{3}~IQ modulator for} coherent data transmission.}
            (a) Optical micrograph of a 13.5~mm-long {\SiN-\LT} IQ modulator. The waveguides are highlighted in red for better visibility. (b)~Schematics of the setup used for the coherent communications experiment. AWG: Arbitrary-waveform generator; Amp.: RF amplifier; Tx-DSP: Transmitter-digital signal processing (offline); ECL: External cavity laser; FPC: Fiber polarization controller; EDFA: Erbium-doped fiber amplifier; BPF: Bandpass filter; TO-PS: Thermo-optic phase shifter; ECL: External-cavity laser; 90$^{\circ}$~OH: 90$^{\circ}$~optical hybrid; BPD: Balanced photodiode; RTO: High-speed real-time oscilloscope (256~GSa/s, 105~GHz); Rx-DSP: Receiver-digital signal processing (offline). (c)~Measured bit error ratio (BER) versus symbol rate for QPSK (brown line) and 16-QAM (dark blue line) signals with horizontal dashed lines indicating the thresholds for soft-decision forward error correction having \SI{15}{\percent} and \SI{25}{\percent} coding overhead (\SI{15}{\percent} SD-FEC, \SI{25}{\percent} SD-FEC) and for hard-decision FEC with \SI{7}{\percent} coding overhead (\SI{7}{\percent} HD-FEC). The gray dashed line at the bottom indicates the detection limit, below which our experiment cannot reliably measure BER due to the limited length of the recorded signals ($2^{23}$~samples). This limit corresponds to 13 detected bit errors at a symbol rate of 160~GBd with QPSK signals during the recording period. The resulting BER falls within the \SI{99}{\percent}~confidence interval, ranging from half to twice the measured BER~\cite{BER_limit}. The constellation diagrams of three selected data points \textcircled{A}, \textcircled{B} and \textcircled{C} are shown in panels (e), (f) and (g). (d)~Achievable information rate (AIR, dashed lines) and corresponding net data rate (NDR, solid lines) versus symbol rate for QPSK (brown lines) and 16-QAM (dark blue lines) signals. (e)-(g)~Constellation diagrams for QPSK signals at symbol rates of 160 GBd (e) and 204~GBd (f), and for the 16-QAM signals at a symbol rate of 176~GBd (g). The corresponding data points are marked by red dashed circles and labeled with \textcircled{A}, \textcircled{B} and \textcircled{C}, respectively, in panel~(c).
                    }
		\label{fig4}
    \end{figure*}
    
    To quantify the information transfer efficiency of our transmission system, we derive the generalized mutual information (GMI) of the transmitted signals from the log-likelihood ratios (LLRs) of the received symbols based on an additive white Gaussian noise (AWGN) channel model \cite{7436797}. Figure~\ref{fig3}(c) shows the resulting achievable information rates (AIR), which correspond to the product of the symbol rate and the GMI of each symbol and provide an upper bound for the transmission capacity of the system. To estimate practically achievable net data rates (NDR), we additionally have to include penalties introduced by typical FEC codes. By comparing the normalized GMI (NGMI) estimated in our experiments to the NGMI thresholds provided in~\cite{Hu2022fec}, we select suitable FEC codes and evaluate the NDR by multiplying the line rates and the associated FEC code rates. The highest NDR of 333~Gbit/s is obtained for a symbol rate of 192~GBd (line rate 384~Gbit/s) in combination with \SI{15}{\percent} SD-FEC, while the AIR at this symbol rate amounts to 349~Gbit/s.

    Besides MZM and IMDD signaling, our {\SiN-\LT} platform also supports more advanced systems, such as IQ modulators (IQM) for coherent communications. Figure~\ref{fig4}(a) shows a micrograph of such an IQM. The device consists of an input waveguide followed by a Y-splitter leading to two 13.5~mm-long MZMs, such as the one shown in \fref{fig2}(a). The individual MZMs are DC-biased at their minimum transmission point using two 110~GHz bias-Tees (BT110R-C, SHF Communication Technologies AG). A heater-based phase shifter, which can produce a drift-free thermo-optic phase shift~\cite{Li:18}, then sets the relative phase between the signals generated in the in-phase (I) and the quadrature (Q) arms of the IQM to $90^\circ$. Thereafter, a power combiner, like the one in \fref{fig2}(a), synthesizes the output signal with two modulated quadratures. To demonstrate the viability of the device, we use the setup shown in \fref{fig4}(b) to transmit quadrature phase-shift keying (QPSK) and 16-state quadrature-amplitude modulation (16QAM) signals at symbol rates between 144~GBd and 204~GBd. The setup shares several similarities with the one used for IMDD transmission shown \fref{fig3}(a). Here, the two AWG channels however individually drive the I and Q arms of the IQM. After modulation, amplification and bandpass filtering, the modulated signal enters a $90^\circ$ optical hybrid module, where it interferes with a continuous-wave local oscillator provided by a second ECL to down-convert the optical signal to four electrical baseband signals. Subsequent balanced photodiodes (BPD, Fraunhofer HHI) then detect the resulting waveforms. Finally, as clarified in Supplementary Section~5, offline receiver DSP recovers and evaluates the signals.
    
    Figure~\ref{fig4}(c) shows the measured BER along with the BER thresholds for different FEC schemes indicated by dashed horizontal lines. Figure~\ref{fig4}(d) provides the corresponding AIR and NDR. For QPSK signals, we obtained reliable BER values only for our measurements at symbol rates of 176~GBd and higher, whereas the measurements at 144~GBd and 160~GBd led to BER values below \num{1.24e-6}, which corresponds to the detection limit for the length of the recorded waveforms. As shown in \fref{fig4}(e), the constellation diagram at 160~GBd features a constellation signal-to-noise ratio (CSNR) of 15.8~dB, which would correspond to an estimated BER of \num{3.5e-10}~\cite[Eq.~2.18]{de_arruda_mello_digital_2021}. At the highest QPSK symbol rate of 204~GBd, we measure a BER of \num{7e-3}, which is still below the BER threshold for SD-FEC schemes with 25~\% overhead. In this case, the CSNR amounts to 8.5~dB as displayed in the corresponding constellation diagram from \fref{fig4}(f). Here, the 75~GHz electrical 3~dB-bandwidth of the AWG~\cite{awg_m8199B} mainly limits the performance of this experiment. As indicated in \fref{fig4}(c), the highest symbol rate achieved for 16-QAM signaling amounts to 176~GBd and leads to a measured BER of \num{3.08e-2} \textendash\ still below the BER threshold for SD-FEC codes with 25~\% overhead. This corresponds to a line rate of 704~Gbit/s. The corresponding NDR of up to 581~Gbit/s validates the feasibility of telecommunication systems based on hybrid {\SiN-\LT} circuits. Figure~\ref{fig4}(g) shows the corresponding constellation diagram, from which we measure a CNSR of 12.0~dB.
    
    To the best of our knowledge, these experiments represent the first demonstration of {\LT-on-\SiN-based} high symbol-rate communication experiments and represents an important milestone that has not been reached for other similar heterogeneous photonic platforms to date~\cite{Li:25, niels2025high,rahman2025high} and exceeds the previously reported symbol rates below $100$~GBd in hybrid silicon lithium niobate coherent modulators~\cite{wang_siliconlithium_2022}.
    

    
    \section{Discussion}

    Alternative {hybrid \SiN-\LT} waveguide geometries can further improve some of their performance figures. However, such improvements might come at the expense of other metrics. For instance, relying on a thinner bonded \LT~film can reduce slab losses in the resulting hybrid waveguide, thereby leading to lower optical propagation losses and ring resonators with higher quality factors. However, this narrower film will also reduce the overlap of the propagating mode with the electro-optic material, thus resulting in lower modulation efficiency. Besides such underlying design tradeoffs, some features might require specific waveguide dimensions that will constrain other figures of merit to fixed ranges. For example, a waveguide hosting optical nonlinearities must often verify specific dispersive relations determined by its cross-section~\cite{Pfeiffer:16, liu_high-yield_2021, ji2025copper}. As a result, distinct features required of the PIC by specific end-uses will ultimately dictate the geometry and hence the performance of its waveguides.

    Our {\SiN-\LT} platform could potentially benefit from additional functionalities brought by platform extensions. Namely, adapting its bonding process could lead to the integration of \threefive active optical materials~\cite{xiang2021high,doi:10.1126/science.abh2076,sun2024high,xie20253} and potentially to a multilayer platform featuring both \threefive components~\cite{xiang20233d} and {\SiN-\LT} hybrid waveguides. Incorporating distinct \SiN~and \LT~waveguide layers would also introduce benefits, such as eliminating slab losses in \SiN~waveguides and increasing mode overlap with \LT~for enhanced electro-optic coupling strengths. However, these benefits will come at the cost of those attributed to the minimal processing of the \LT~film in our fabrication flow.
    
    
    To summarize, we introduced a photonic platform relying on wafer-scale bonding of \LT~films on foundry-compatible \SiN~PICs. Its fabrication process reaches a 100\% bonding yield across the wafer's nine central device fields. Compared to other monolithic electro-optic PIC platforms~\cite{wang_ultrabroadband_2024,xu2020high,han2023slow}, {\SiN-\LT} circumvents specialized \LT~etching by relying on a standardized \SiN~high-volume process flow. Resulting circuits preserve low optical losses at telecommunication wavelengths  while also accommodating broadband modulators featuring high efficiencies and a flat electro-optic response over modulation rates extending up to 100~GHz. Data transmission experiments confirm the practicality of these features by demonstrating net data rates exceeding 500~Gbit/s. Ease of access to high-volume production of {\SiN-\LT} PICs reinforces its prospects in field-deployable applications not only limited to high-speed optical communications, but that also include microwave-to-optical transducers~\cite{holzgrafe2020cavity,shen2024photonic,warner2025coherent} and fast tunable LiDAR \cite{li2023frequency,siddharth2025ultrafast}. Built-in integration with thick \SiN waveguides also enables a new generation of on-chip systems leveraging both electro-optic and ultralow loss waveguides ranging from microwave oscillators~\cite{kudelin_photonic_2024, He:24}, to potential interfaces with Kerr frequency combs~\cite{liu_high-yield_2021, ji_compact_2022}.

	\begin{footnotesize}
		
		\noindent \textbf{Note:}  During the preparation of this manuscript, two similar manuscripts \cite{niels2025high, rahman2025high} were posted on the arXiv pre-print repository, but with distinct fabrication processes and experimental results. 

		\noindent \textbf{Author Contributions}: J.C., C.W., and X.J. fabricated the \SiN-\LT~PICs. J.C., C.W, and J.Z. measured the losses and EO responses of the PICs. A.K. measured the frequency response. A.K. and D.D. conceived and performed the data transmission experiments together with C.K. and jointly discussed the results. J.C., A.K., D.D., and H.L. analyzed the data. T.J.K., C.K., X.O. and H.L. supervised all aspects of the project. H.L. and J.C. prepared the content of the manuscript in assistance with contributions and discussions provided by all the authors.

		\noindent \textbf{Funding Information and Disclaimer}: This work was supported by funding from the Swiss National Science Foundation under grant agreement No. 216493 (HEROIC)), by funding from the German Research Foundation via the projects PACE (\# 403188360) and GOSPEL (\# 403187440), and by the European Innovation Council (EIC) via the project ELLIPTIC (\# 101187515).

		\noindent \textbf{Acknowledgments}:
		The PICs were fabricated in the EPFL Center of MicroNanoTechnology (CMi). The LTOI wafers were fabricated in Shanghai Novel Si Integration Technology (NSIT) and the SIMIT-CAS.

            \noindent \textbf{Competing Interests}:
		C.K. and T.J.K are co-founders and shareholders of Luxtelligence SA, St. Sulpice, Switzerland, a company engaged in electro-optic modulators based on ferroelectric materials.
		
		\noindent \textbf{Data and Code Availability Statement}: The code and data used to produce the plots within this work will be released on the repository \texttt{Zenodo} upon publication of this preprint.

	\end{footnotesize}
       
        \bibliography{citations}

\end{document}


\title{Supplementary Information for: Heterogeneously integrated lithium tantalate-on-silicon nitride modulators for high-speed communications}

\author{Jiachen~Cai}\thanks{These authors contributed equally.}
    \affiliation{State Key Laboratory of Materials for Integrated Circuits, Shanghai Institute of Microsystem and Information Technology, Chinese Academy of Sciences, Shanghai, China}
    \affiliation{Institute of Physics, Swiss Federal Institute of Technology Lausanne (EPFL), CH-1015 Lausanne, Switzerland}

    \author{Alexander~Kotz}\thanks{These authors contributed equally.}
    \affiliation{Institute of Photonics and Quantum Electronics (IPQ), Karlsruhe Institute of Technology (KIT), 76131 Karlsruhe, Germany}

    \author{Hugo~Larocque}\thanks{These authors contributed equally.}
    \affiliation{Institute of Physics, Swiss Federal Institute of Technology Lausanne (EPFL), CH-1015 Lausanne, Switzerland}
    
    \author{Chengli~Wang}
    \affiliation{State Key Laboratory of Materials for Integrated Circuits, Shanghai Institute of Microsystem and Information Technology, Chinese Academy of Sciences, Shanghai, China}
    \affiliation{Institute of Physics, Swiss Federal Institute of Technology Lausanne (EPFL), CH-1015 Lausanne, Switzerland}

    \author{Xinru~Ji}
    \affiliation{Institute of Physics, Swiss Federal Institute of Technology Lausanne (EPFL), CH-1015 Lausanne, Switzerland}

    \author{Junyin~Zhang}
    \affiliation{Institute of Physics, Swiss Federal Institute of Technology Lausanne (EPFL), CH-1015 Lausanne, Switzerland}

    \author{Daniel~Drayss}
    \affiliation{Institute of Photonics and Quantum Electronics (IPQ), Karlsruhe Institute of Technology (KIT), 76131 Karlsruhe, Germany}

    \author{Xin~Ou}
    \email[]{ouxin@mail.sim.ac.cn}
    \affiliation{State Key Laboratory of Materials for Integrated Circuits, Shanghai Institute of Microsystem and Information Technology, Chinese Academy of Sciences, Shanghai, China}

    \author{Christian~Koos}
    \email[]{christian.koos@kit.edu}
    \affiliation{Institute of Photonics and Quantum Electronics (IPQ), Karlsruhe Institute of Technology (KIT), 76131 Karlsruhe, Germany}
    
    \author{Tobias~J.~Kippenberg}
    \email[]{tobias.kippenberg@epfl.ch}
    \affiliation{Institute of Physics, Swiss Federal Institute of Technology Lausanne (EPFL), CH-1015 Lausanne, Switzerland}
    \affiliation{Institute of Electrical and Micro engineering, Swiss Federal Institute of Technology, Lausanne (EPFL), CH-1015 Lausanne, Switzerland}

\setcounter{equation}{0}
\setcounter{figure}{0}
\setcounter{table}{0}

\setcounter{subsection}{0}
\setcounter{section}{0}
\setcounter{secnumdepth}{3}

\maketitle
{\hypersetup{linkcolor=blue}\tableofcontents}
\newpage

\section{Photonic Damascene process for integrated silicon nitride waveguides}
\label{sec:fab}

\begin{figure*}[htbp!]
	\centering
	\includegraphics[width=1\linewidth]{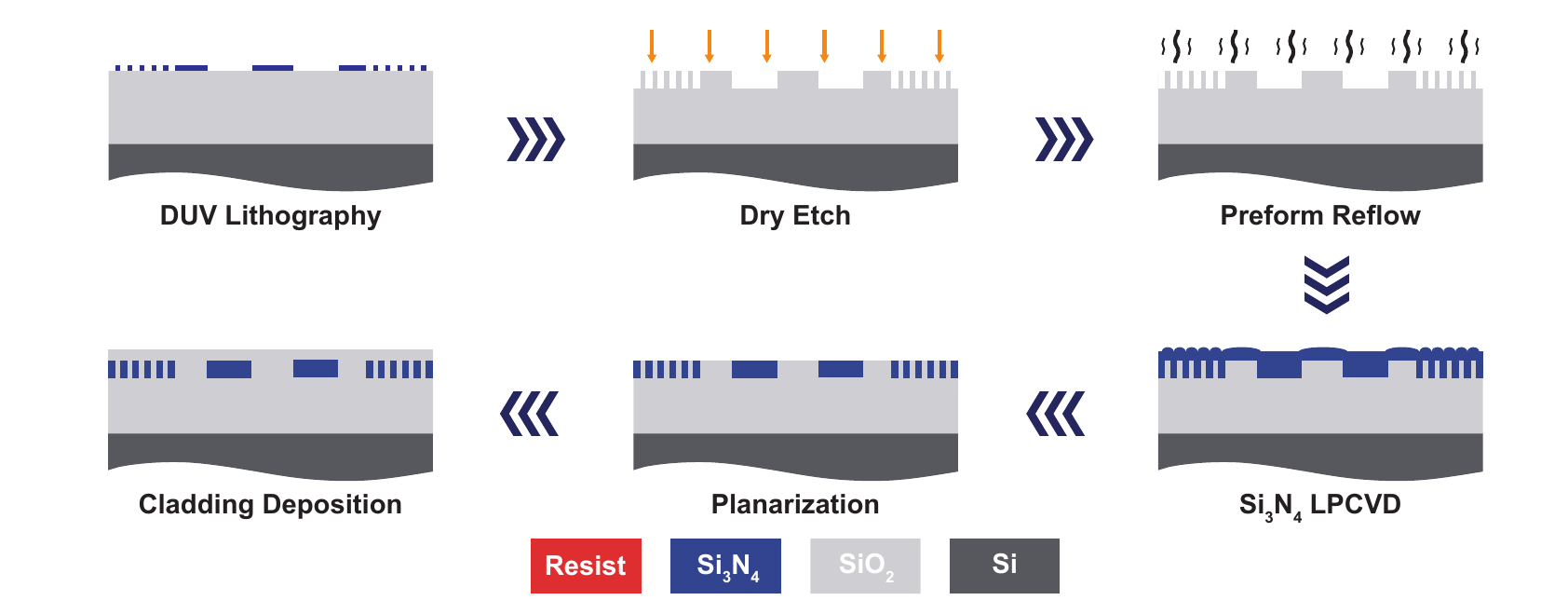}
	\caption{\textbf{Wafer-scale photonic Damascene process for low-loss \SiN photonic integrated circuits.}
        The process flow includes deep ultraviolet (DUV) stepper lithography, fluoride dry etching, preform reflow, high-density \SiN deposition, chemical mechanical polishing (CMP), and cladding deposition.
		}
	\label{figS1}
\end{figure*}

\indent The process begins with a standard wet oxide substrate (\SI{4}{\micro\meter}~\SiO/\SI{525}{\micro\meter} Silicon). The waveguide and filler patterns are defined on the top silica using an ASML PAS 5500/350C DUV stepper with a 248 nm wavelength light source. The micron-order filler patterns are designed as quasi-intersecting line structures to mitigate the tensile stress associated with high density \SiN~deposition. These patterns are then transferred into the underlying wet oxide via fluorine-based dry etching to a depth of 500~nm. A subsequent high-temperature annealing step at 1250$^\circ$C is performed to promote oxide reflow and improve sidewall smoothness. After annealing, a 700 nm-thick \SiN~layer is deposited into the trenches using~LPCVD. A dedicated-adjusted CMP process is implemented to remove the redundant material outside the trench, as well as enabling ultralow surface roughness on the top surface of \SiN~waveguides. A small amount of oxide interlayer is then deposited by an inductively-coupled-plasma CVD (ICPCVD) with $\mathrm{SiCl}_4$ gas as the precursor. An additional 1200$^\circ$C annealing is conducted to eliminate absorption losses associated with hydrogen impurities. Finally, a second CMP step is adopted to ensure sub-nanometer surface roughness for subsequent wafer bonding of lithium tantalate on Damascene silicon nitride wafers.

\section{Simulations for silicon nitride-lithium tantalate Mach-Zehnder modulators}

\subsection{Trade-off between $V_{\pi}\cdot L$ and optical loss}
\begin{figure*}[htbp!]
	\centering
	\includegraphics[width=1\linewidth]{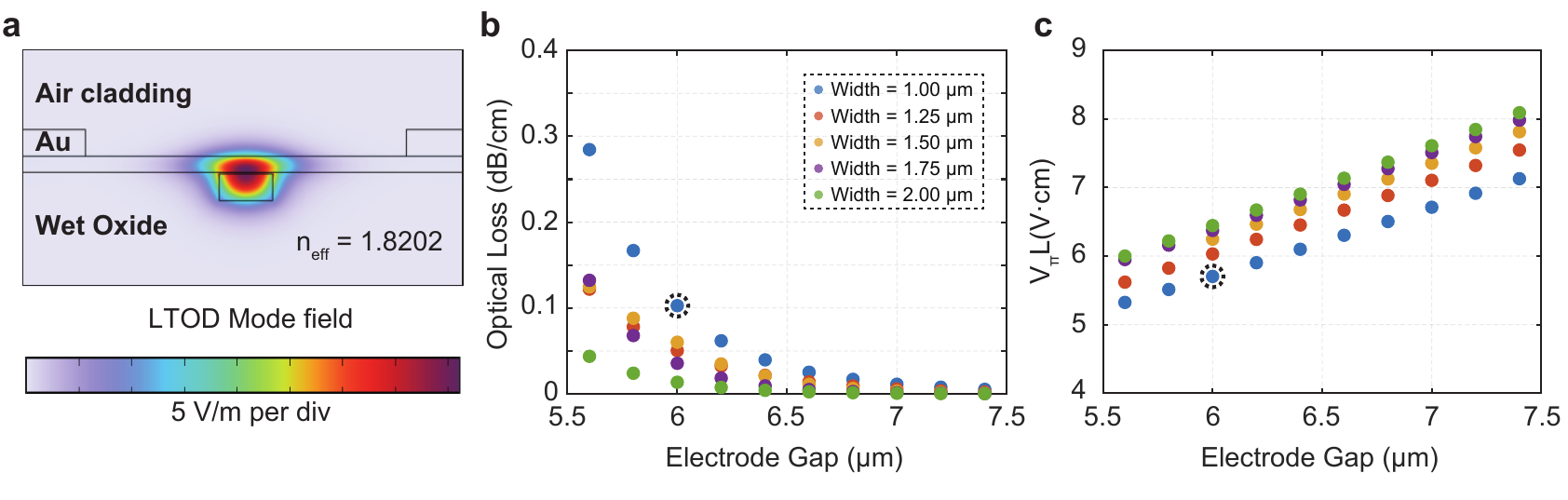}
	\caption{\textbf{Static field simulation of silicon nitride-lithium tantalate Mach-Zehnder modulators.}
	(a) Simulated optical mode field in the cross-section of the hybrid \LT-\SiN waveguide. (b) Calculated optical loss and (c) $V_{\pi}\cdot L$ versus different electrode gap values and \SiN waveguide widths. The hollow circles mark the estimated performance for the fabricated {\SiN-\LT} modulators reported in this work, with a waveguide width of \SI{1}{\micro\meter} and an electrode gap of \SI{6}{\micro\meter}.}
	\label{fig2}
\end{figure*}

Optical simulations (COMSOL multiphysics) were first employed to calculate the mode profiles in {\SiN-\LT} hybrid waveguides. Given the fabrication process from the main text and Section~\ref{sec:fab}, the thickness of \LT and \SiN is chosen to be 300 nm and 500 nm, respectively. Figure \ref{fig2}(a) represents the electric field distribution of the hybrid optical mode. The confined mode is divided roughly equally between the \SiN ridge waveguide and the lithium tantalate thin film. 
As stated in the Discussion part of the main text, alternative mode distributions are feasible by altering the film thickness, but it involves changes in the preparation of the lithium tantalate wafer, such as adjusting the ion-implantation dose and annealing temperature during the corresponding smart-cut process~\cite{9933105,FENG20044299}. Furthermore, modifying the film thickness also introduces a performance trade-off between actuation voltage and optical insertion loss. Efficient electro-optic (EO) modulation can be achieved by adopting a close enough electrode-to-waveguide spacing at the expense of larger optical absorption. Here, the device geometry is well optimized using finite element simulations. Assuming that the propagation mode is transverse electric (TE), the estimated loss related to various electrode gaps can be extracted from the imaginary part of the effective mode index ($n_{eff}$),
\begin{equation}
	\alpha = 0.1 \times \vert \log_{10}(e^{-\frac{4\pi}\lambda\cdot \mathrm{Im}(n_{eff})})\vert~~ (dB/cm),
    \label{eq:loss_cal}
\end{equation}
where $\lambda$ represents the wavelength of the selected mode. 
Figure \ref{fig2}(b) shows how the electrode gap affects transmission loss for various \SiN waveguide widths.
The resulting voltage-length products ($V_{\pi}\cdot L$) for the push-pull configuration can be expressed as \cite{Liu:21}:
\begin{equation}
	V_{\pi}\cdot L = \frac{1}{2} \frac{\lambda}{n^{3}r_{33}\Gamma},
    \label{eq:VpiL}
\end{equation}
where $n$ is the refractive index of lithium tantalate and $r_{33}$ is the EO coefficient in the crystallographic c-axis. The mode overlap $\Gamma$ can be given by 
\begin{equation}
	\Gamma = \frac{\int\int \frac{E_{z}(x,z)}{V}\cdot \vert e_{z}(x,z) \vert^2 dxdz}{\int\int \vert e_{z}(x,z) \vert^2 dxdz},
\end{equation}
where $e_z(x,z)$ and $E_z(x,z)$ represent the horizontal electric field components of the optical TE mode and actuation field from the electrodes, respectively.
As plotted in Fig.~\ref{fig2}(c), the \SI{1}{\micro\meter}~-wide waveguide shows higher modulation efficiency because of a larger optical mode overlap with the lithium tantalate thin film. Based on the above analysis, we opt for a device geometry featuring a signal-ground spacing of \SI{6}{\micro\meter} and a \SiN waveguide width of \SI{1}{\micro\meter} to obtain a high modulation efficiency with negligible insertion loss.

\subsection{Wave velocity matching}

\begin{figure*}[htbp!]
	\centering
	\includegraphics[width=0.9\linewidth]{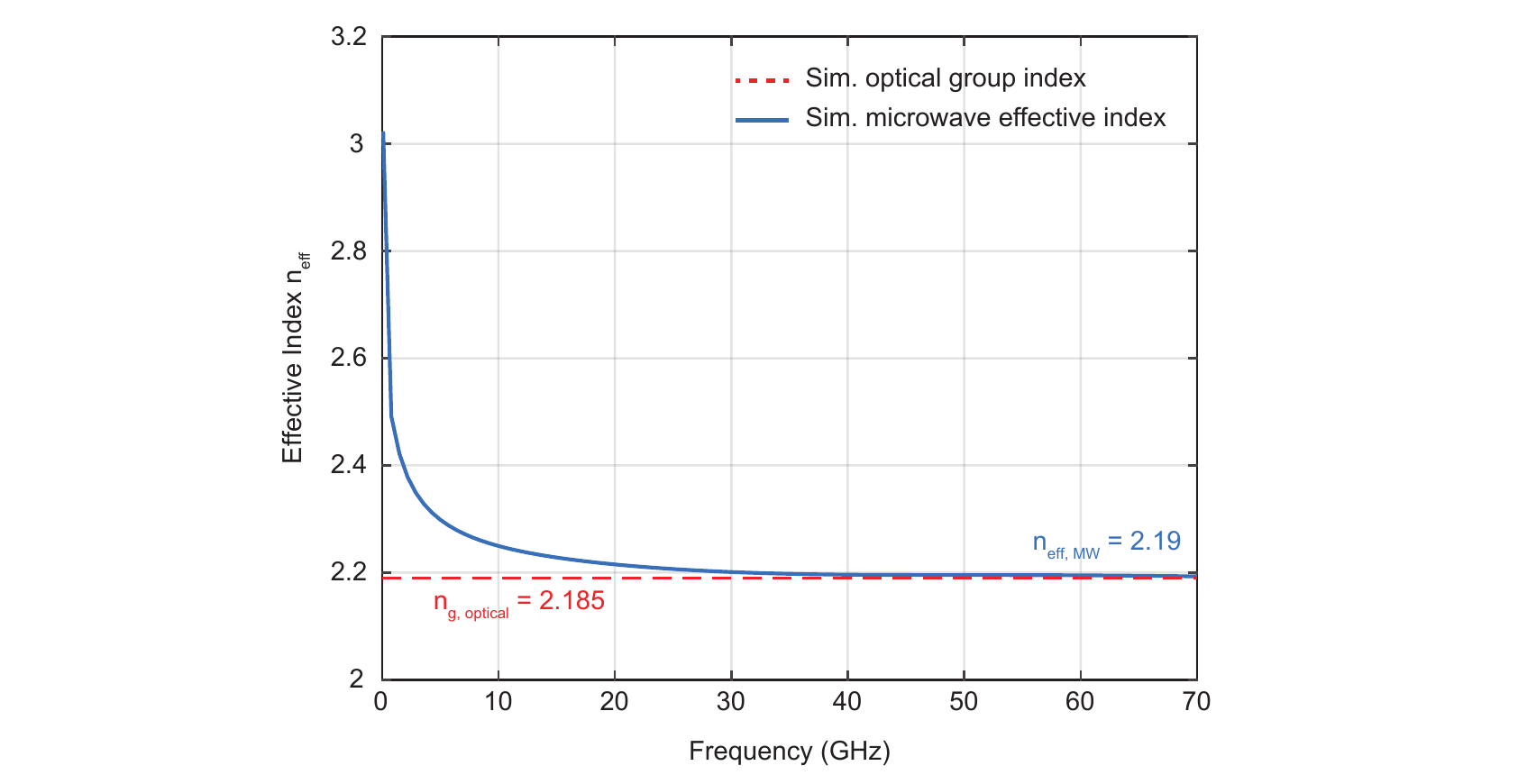}
	\caption{\textbf{Velocity mismatch between microwave and lightwave.}
	Simulated optical group index and microwave effective index, showing a negligible disparity in the high modulation frequency range.
		}
	\label{fig3}
\end{figure*}

To sustain interactions between propagating microwaves and optical modes over long distances, the velocities of these two fields must be similar~\cite{hu2025integrated}. For a fixed optical waveguide geometry, such a requirement can be satisfied by a suitable traveling wave electrode design. We simulate the effective refractive index ($n_{eff,MW}$) of a propagating RF field in high-speed electrodes using Ansys HFSS. The $n_{eff,MW}$ can be extracted by phase unwrapping the transmission $S_{21}$ of coplanar waveguide (CPW) electrodes:

\begin{equation}
	n_{eff} = c_{0}\frac{\mathrm{unwrap}(ang(S_{21}))}{2\pi \omega_{MW} L_{ele}},
\end{equation}
where $c_0$ is the speed of light in vacuum, $\omega_{RF}$ is the RF modulation frequency and $L_{ele}$ is the length of the CPW electrodes. The group refractive index ($n_{g,optical}$) of the waveguide's optical TE mode is obtained using Ansys Lumerical Mode.
Figure \ref{fig3} implies a near-perfect alignment between the optical $n_g$ and the microwave $n_{eff}$ above a modulation frequency of 40 GHz, thereby verifying the required phase-matching behavior for an {\SiN-\LT} modulator with a high EO bandwidth.

\section{Simulations for tapered mode transitions}

\begin{figure*}[htbp!]
	\centering
	\includegraphics[width=0.9\linewidth]{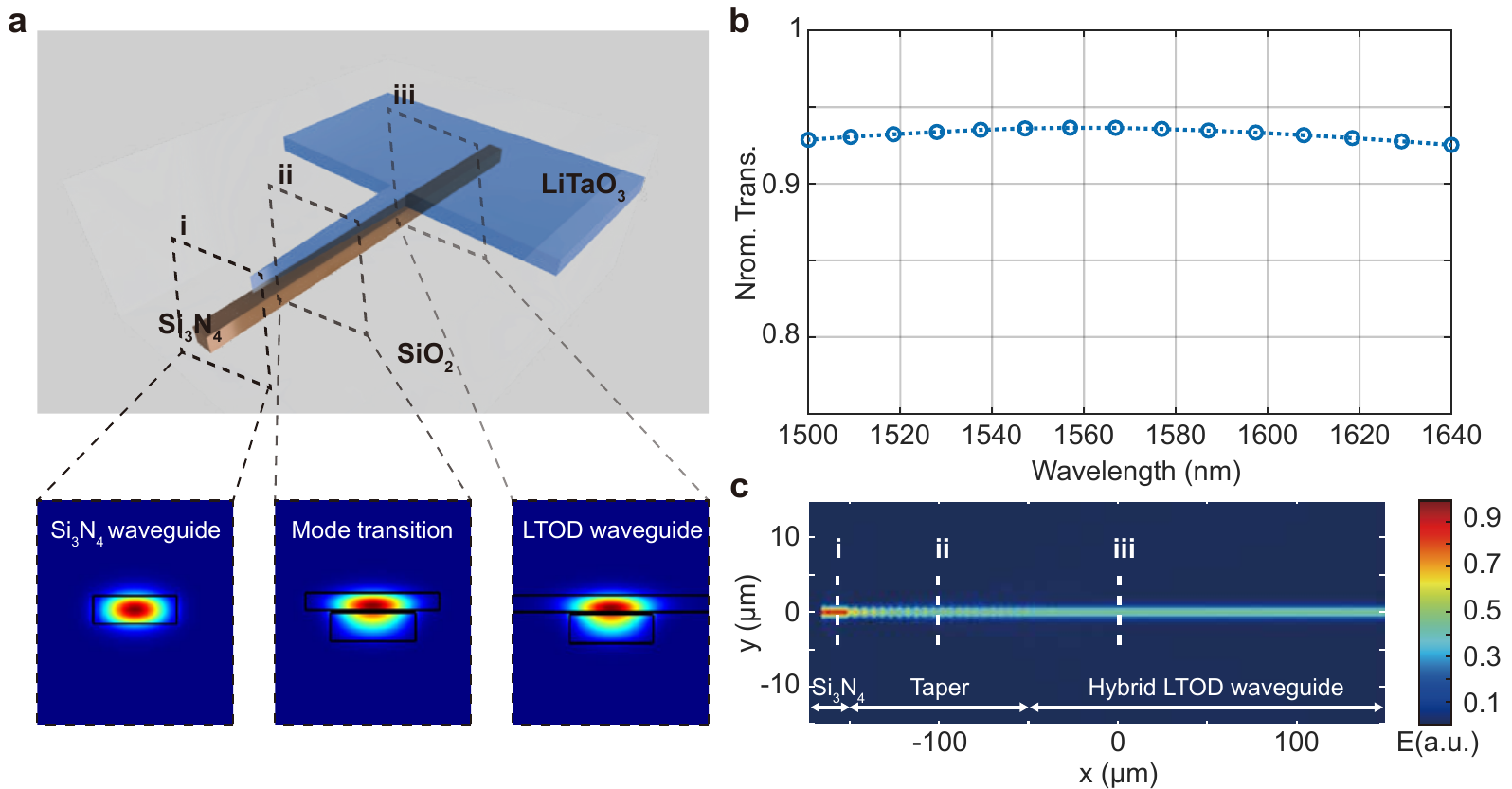}
	\caption{\textbf{Simulated transmission of silicon nitride-lithium tantalate waveguide transitions.} (a) Schematic diagram and corresponding FDTD simulations of the adiabatic coupling from the \SiN waveguide to the hybrid \SiN-\LT waveguide. Simulated (b) transmission spectrum and (c) electric field distribution of the adiabatic taper. 
	}
	\label{fig4}
\end{figure*}

Low-loss optical mode coupling is achieved using an adiabatic taper transition \cite{churaev2023heterogeneously}, enabling efficient single-mode transition from \SiN waveguides to hybrid {\SiN-\LT} waveguides (Fig.~\ref{fig4}(a)). Based on the fabricated \SiN-\LT structure, the coupling design exhibits a \SI{100}{\micro\meter}-long inverse taper and a \SI{500}{\nano\meter}-wide tip, which are constructed in the etched \LT film. As illustrated in Figs.~ \ref{fig4}(b,c), this tapered design facilitates high coupling efficiency across a broad wavelength range (\SI{1500}{\nano\meter} to \SI{1640}{\nano\meter}) for {\SiN-\LT} modulators, achieving a low insertion loss of near 0.28~dB per facet.

\section{Microwave transmission and modulation efficiency}

\begin{figure*}[htbp!]
	\centering
	\includegraphics[width=0.9\linewidth]{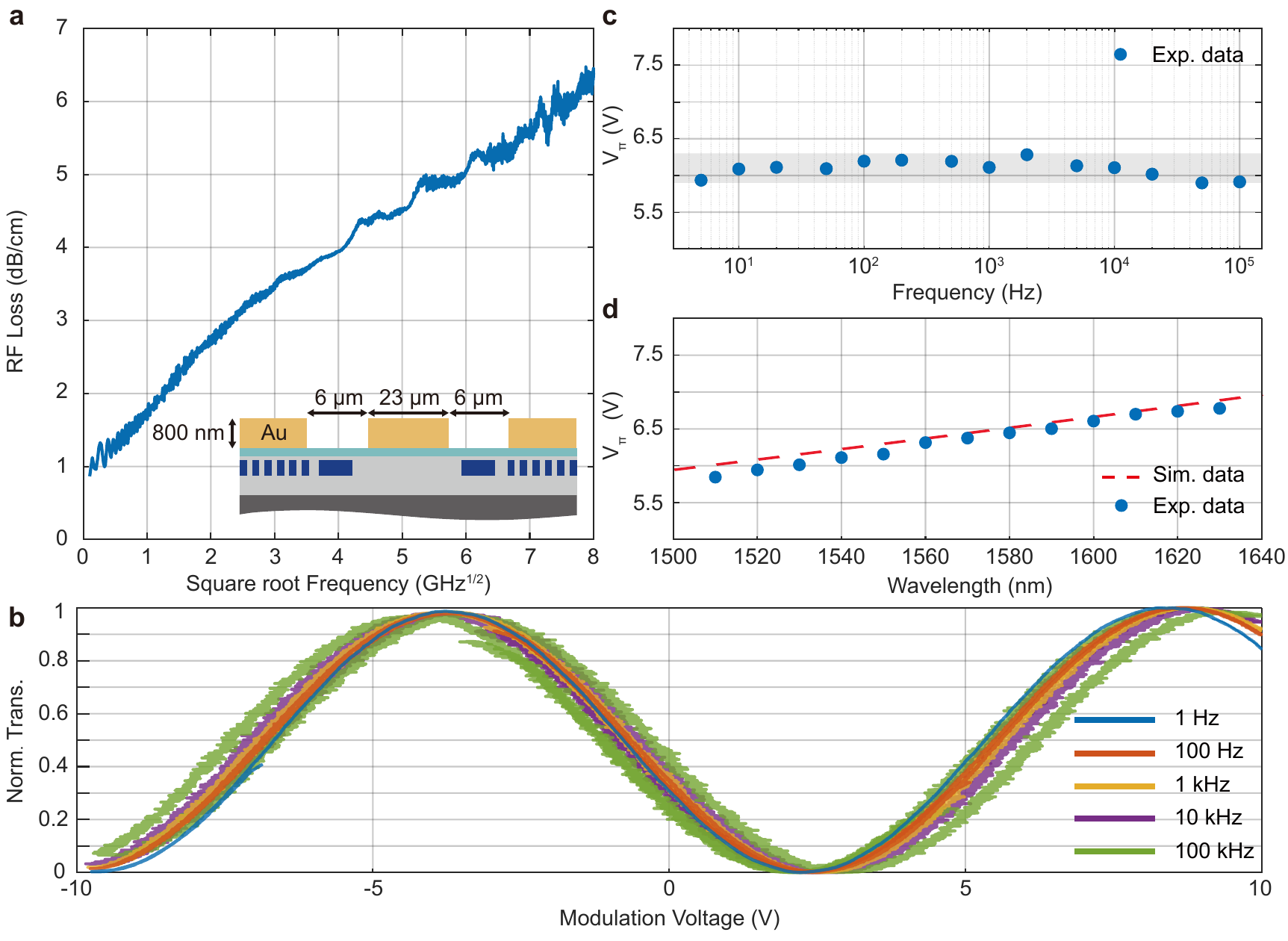}
	\caption{\textbf{Modulator microwave transmission and modulation efficiency.} (a) Measured co-planar waveguide microwave losses on a square-root frequency axis. (b) Measured transmission of the {\SiN-\LT} modulators driven by modulation frequencies from 1 Hz to 100 kHz. Extracted modulator $V_{\pi}$ value at (c) different modulation frequencies at a fixed optical \SI{1550}{\nano\meter} optical wavelength and (d) different optical wavelengths at a fixed \SI{100}{\hertz} modulation frequency.
    }
	\label{fig5}
\end{figure*}

\indent We first characterized the RF attenuation properties of the fabricated CPW-type modulators. Figure \ref{fig5}(a) presents the extracted microwave loss of the long gold electrode on the {\SiN-\LT} platform, using a 67 GHz vector network analyzer (VNA). The loss profile demonstrates performance comparable to state-of-the-art ultrabroadband EO modulators reported in literature \cite{wang_ultrabroadband_2024,xu2020high}. 
Additionally, the square-root frequency dependence of RF loss in figure \ref{fig5}(a) indicates the dominant ohmic loss mechanism ($\alpha\propto \sqrt{f_{MW}}$), confirming that our fabrication techniques can effectively prevent the devices from parasitic-capacitance-induced loss \cite{shen_parasitic_2024,photonics11050399}.

To further demonstrate the dynamic performance of {\SiN-\LT} Mach-Zehnder modulators, we then investigated their EO performance under low-frequency RF modulation ranging from 1 Hz to 100 kHz (Fig.~\ref{fig5}(b)). Despite the influence of ferroelectric hysteresis, we extracted a stable half-wave voltage ($V_{\pi}$) with an average $V_{\pi}$ value of 6.1 V (Fig.~\ref{fig5}(c)) from the experimental data. 
Similarly, we studied the wavelength response of the modulator performance driven by a 100 Hz triangle wave signal (Fig.~\ref{fig5}(d)), which features excellent agreement with our simulation results (red dashed line). This wavelength-dependent behavior is due to the increased optical mode expansion in the lithium tantalate thin film at longer wavelengths, leading to an enhanced microwave-optical field interaction.

\section{Digital signal processing at the receiver for silicon nitride-lithium tantalate modulator-based communication experiments}
\indent In the data transmission experiments discussed in the main manuscript, the optical signals were detected by photodiodes and the resulting photocurrents were digitized by a high-speed real-time oscilloscope (UXR 1004A, Keysight Technologies Inc.) operating at a sampling rate of 256~GSa/s with an analog bandwidth of 105~GHz. A total of $2^{23}$~samples corresponding to a time interval of approximately \SI{33}{\micro\second} were recorded. Offline, non-data-aided signal processing was employed to extract the transmitted data.

For both IMDD and coherent transmission schemes, the digitized signal was initially resampled to two samples per symbol. Timing recovery was then performed using a feedforward timing recovery algorithm, as described in~\cite{barton_symbol_1992} and~\cite{matalla_hardware_2021}, to estimate and correct the timing offset of the received signal. An adaptive receive filter, implemented as a time-domain linear equalizer, was subsequently applied.

In the IMDD case, linear Sato equalization~\cite{sato_method_1975} was utilized, followed by a linear post-equalizer based on the decision-directed least-mean-squares (DD-LMS) algorithm~\cite{randel_all-electronic_2014} to recover the transmitted data.

For coherent communications, a linear equalizer based on the constant modulus algorithm~\cite{moshirian_blind_2012} was applied. This was followed by frequency offset compensation using a phase increment estimation algorithm~\cite{leven_frequency_2007}, which corrects for the frequency mismatch between the optical carrier at the transmitter and the local oscillator at the receiver. Residual phase errors, originating from the phase noise of the transmitter and receiver lasers, was mitigated using the blind phase search algorithm~\cite{pfau_hardware-efficient_2009}. Finally, a DD-LMS-based linear post-equalizer~\cite{randel_all-electronic_2014} was employed to recover the complex-valued transmitted symbols.

\bibliography{SIbib}